\documentclass[%
preprint,
showpacs,preprintnumbers,
 amsmath,amssymb,
 aps,
prb,
floatfix,
]{revtex4-1}

\usepackage{graphicx}
\usepackage{dcolumn}
\usepackage{bm}

\usepackage{pstricks}
\newrgbcolor{redtwo}{.95 .05 0}

\begin{document}

\preprint{APS/123-QED}

\title{Electron localization and possible phase separation in the absence of a charge density wave in single-phase 1T-VS$_2$}

\author{A. Gauzzi}
\email{andrea.gauzzi@upmc.fr}
\author{A. Sellam}
\author{G. Rousse}
\author{Y. Klein}
\author{D. Taverna}
\author{P. Giura}
\author{M. Calandra}
\author{G. Loupias}
\affiliation{IMPMC, UPMC-Sorbonne Universit\'es, CNRS, MNHN, IRD, 4, place Jussieu, 75005 Paris, France}

\author{F. Gozzo}
\thanks{Present address: Excelsus Structural Solutions sprl, 1150 Bruxelles, Belgium.}
\affiliation{Swiss Light Source, Paul Scherrer Institut, 5232 Villigen, Switzerland}

\author{E. Gilioli}
\author{F. Bolzoni}

\affiliation{Istituto dei materiali per elettronica e magnetismo\\
Consiglio Nazionale delle Ricerche, Area delle Scienze, 43100 Parma, Italy}

\author{G. Allodi}
\author{R. De Renzi}
\affiliation{Dipartimento di fisica, Universit\`a di Parma, 43100 Parma, Italy}

\author{G. L. Calestani}
\affiliation{Dipartimento di chimica - GIAF, Universit\`a di Parma, 43100 Parma, Italy}

\author{P. Roy}
\affiliation{Synchrotron Soleil, L'Orme des M\'erisiers, Saint-Aubin - 91192 Gif-sur-Yvette, France}

\date{\today}

\begin{abstract}
We report on a systematic study of the structural, magnetic and transport properties of high-purity 1T-VS$_2$ powder samples prepared under high pressure. The results differ notably from those previously obtained by de-intercalating Li from LiVS$_2$. First, no Charge Density Wave (CDW) is found by transmission electron microscopy down to 94 K. Though, \textit{ab initio} phonon calculations unveil a latent CDW instability driven by an acoustic phonon softening at the wave vector ${\bf q}_{CDW} \approx$ (0.21,0.21,0) previously reported in de-intercalated samples. A further indication of latent lattice instability is given by an anomalous expansion of the V-S bond distance at low temperature. Second, infrared optical absorption and electrical resistivity measurements give evidence of non metallic properties, consistent with the observation of no CDW phase. On the other hand, magnetic susceptibility and NMR data suggest the coexistence of localized moments with metallic carriers, in agreement with \textit{ab initio} band structure calculations. This discrepancy is reconciled by a picture of electron localization induced by disorder or electronic correlations leading to a phase separation of metallic and non-metallic domains in the nm scale. We conclude that 1T-VS$_2$ is at the verge of a CDW transition and suggest that residual electronic doping in Li de-intercalated samples stabilizes a uniform CDW phase with metallic properties.

\end{abstract}



\maketitle

\section{\label{intro}Introduction}
There has recently been a renewed interest in the interplay between Charge Density Wave (CDW) and superconducting states in layered dichalcogenides $MX_2$ ($M$=transition metal; $X$=S,Se), such as TiSe$_2$, TaSe$_2$ or TaS$_2$. The CDW mechanism remains controversial and various scenarios beyond the classic mechanism of Peierls instability \cite{pei55}, such as phonon softening \cite{cal11}, Coulomb repulsion \cite{faz80,ros06,sip08} or exciton condensation \cite{cer07} have been invoked. Clarifying this controversy may help to explore the possibility of unconventional, \textit{e.g.} exciton-mediated, superconductivity in this system, as suggested by Ginzburg for metal-dielectric bilayers \cite{gin70}.  

In order to address this issue, we carried out a systematic study of the structural, magnetic and transport properties of pure VS$_2$, a model $d^1$ system with quasi two-dimensional properties. Its 1T (or CdI$_2$-type) structure (see Fig.~\ref{fig:XRD}) is made of layers of VS$_6$ octahedra separated by a van der Waals gap and is described by the $P\overline 3m$1 symmetry. As compared to the isoelectronic and isostructural compound 1T-TaS$_2$, which has been extensively studied, VS$_2$ is simpler owing to the absence of superconductivity and to the lower $Z$-value of V, which leads to a small spin-orbit coupling. Hence, VS$_2$ is a model system to study the stability conditions of the CDW phase. Despite these favorable characteristics, the literature on VS$_2$ is limited because of its metastability \cite{mur77} and V-rich V$_{1+x}$S$_2$ phases, where interstitial V atoms are located between the layers, are obtained at ambient pressure \cite{kat79,pod02}. The pure ($x$=0) phase has been hitherto synthesized only by de-intercalating Li from LiVS$_2$ \cite{mur77,mul10}. Early magnetic \cite{mur77} and NMR \cite{tsu83} studies on such de-intercalated samples point at a CDW phase below 305 K, however this phase was observed only recently by transmission electron microscopy \cite{mul10} which shows an incommensurate in-plane propagation vector, ${\bf q}_{CDW}$=(0.21,0.21,0). Similarly to 1T-TaS$_2$ at high temperature or to 2H-TaS$_2$ at low temperature, the incommensurate CDW phase of VS$_2$ is concomitant to metallic properties. Indeed, only commensurate CDW phases are insulating, as in TiSe$_2$ and TaS$_2$ at low temperature. Our main result is that no CDW metallic phase is found in high-purity VS$_2$ samples synthesized under high-pressure.

The paper is organized as follows. In section II, we report on the sample preparation under high pressure and on the experimental methods. In section III, we report on the main experimental results, the salient results being the evidence of the absence of the CDW phase concomitant to the observation of electron localization. Section IV is devoted to the discussion of the results and to the conclusions. 

\section{\label{exp}Experimental}
\subsection{\label{HP}High-pressure synthesis}
We reproducibly synthesized pure VS$_2$ powders under high pressure using a multi-anvil apparatus. We first prepared a 1:2.05 mixture of vanadium metal and sulfur powders within a high-pressure capsule made of thin Pt foil. A 5 \% sulfur excess was used to ensure the full oxidation of V. The capsule was subsequently kept at 5 GPa and 700 $^{\circ}$C for two hours. Preliminary measurements using a commercial x-ray diffractometer equipped with a standard Cu K$_{\alpha}$ source indicate that the as-prepared powders are single-phase within the sensitivity limit of the apparatus. The external part of the capsule in contact with the Pt foil exhibits the presence of PtS$_2$ and was removed. The single-phase stoichiometric properties of the powders were confirmed by a subsequent synchrotron x-ray diffraction study described below. Our strategy of using high pressure synthesis has been motivated by two early studies showing that, by means of high pressure synthesis, the excess of vanadium is reduced to $x$=0.18 at 0.2 GPa \cite{yok85} and to $x$=0.11 at 2 GPa \cite{nak76}. An extrapolation of these results to higher pressures suggests that the $x$=0 phase should be stabilized at 5 GPa, which is confirmed by our present finding.

\subsection{\label{struct}Structural study}
The as-prepared powders have been studied by means of synchrotron x-ray powder diffraction using a wavelength   $\lambda=0.49575$ \AA~ at the Materials Science beamline of the Swiss Light Source at the Paul Scherrer Institut. The characteristics of the beamline are described in detail elsewhere \cite{pat05}. Salient feature is the use of the high-resolution and fast Mythen microstrip detector enabling the simultaneous detection of the diffracted intensity over $120^{\circ}$ in 2$\vartheta$ in the Debye-Scherrer (transmission) geometry. The substantially reduced data acquisition time and the parallel detection allowed us to perform an accurate structural study as a function of temperature in a wide 5-270 K range. In order to avoid preferential orientation effects, which are expected considering the layered structure of VS$_2$, the powders were placed in a spinning glass capillary mounted on the sample holder. The data were collected during both the cooling-down and the heating-up of the samples in order to exclude thermal hysteresis effects. The structural properties were further investigated using a JEOL 2100F high resolution transmission electron microscope (HRTEM) reaching 1.9 \AA~ resolution and equipped with a 200 keV field-emission electron gun. The measurements were carried out on individual $\sim$ 1 $\mu$m size crystallites at room and liquid nitrogen temperatures. In the latter case, the temperature of the sample holder was $\approx$94 K. No alteration of the structure induced by the electron beam was detected.

\subsection{Magnetic and transport properties}
The dc magnetization of the as-prepared powders was measured in the 5-300 K range using a Quantum Design SQUID magnetometer equipped with a 5 T NbTi superconducting magnet. Both, zero-field-cooling and field cooling curves were taken at 10 oersted. Magnetization curves were taken also as a function of field up to 5 T at fixed temperatures in the 5-300 K range. In order to probe selectively the magnetic behavior of the V ions, $^{51}V$ NMR experiments were performed by employing a home-build spectrometer \cite{all05} and a variable-field superconducting magnet. Spectra were recorded in field sweep mode at a fixed frequency of 78.3 MHz by employing a standard $90^{\circ} - \tau - 90^{\circ}$ spin-echo pulse sequence with a delay, $\tau$, of 10 $\mu$s and a pulse duration of 2.4 $\mu$s. Spin lattice relaxation were measured by the saturation recovery method.

The frequency-dependent optical absorption, $A_{\omega}$, was measured at the AILES beamline of the SOLEIL synchrotron facility on a pellet obtained by intimately mixing 0.5 \% weight of VS$_2$ powders with CsI powders. The characteristics of this beamline are described in detail elsewhere \cite{roy06}; in summary, we used a pulse tube closed cycle cryostat by Cryomech to vary the temperature within the 4-300 K range and a commercial Bruker IFS125/HR spectrometer to measure the transmission in the far infrared 150-650 cm$^{-1}$ range. A spectral resolution of 1 cm$^{-1}$ was achieved using a 6 $\mu$m beamsplitter combined with a commercial Infrared Lab bolometer detector. The absorption of VS$_2$ was obtained by subtracting the absorption of a pure CsI pellet from the absorption of the VS$_2$-CsI pellet; the absorbance was then determined using the usual relation. The study of the transport properties of VS$_2$ was complemented by a dc electrical resistivity measurement down to 2 K on a as-prepared sintered sample using a commercial Quantum Design Physical Property Measurement System.

\subsection{Band structure calculations}
First principles calculations were performed using density functional theory (DFT) in the linear response\cite{RevModPhys.73.515} using the Quantum-Espresso code \cite{gia09} within the local density approximation (LDA) \cite{per81}. We used norm-conserving \cite{PhysRevB.43.1993} and ultrasoft \cite{PhysRevB.41.7892} pseudopotentials for S and V, respectively, and included semicore states in the V pseudopotential. Energy cut-off values were $60$ Rydberg for the kinetic energy expansion and $600$ Rydberg for the charge density. The Brillouin zone integration was performed over a $24\times 24\times 12$ electron-momentum grid and a $4\times 4\times 2$ phonon-momentum grid. The Fermi surface Hermitean-Gaussian smearing used in the simulation was $0.01$ Rydberg.

\section{\label{res}Results}

\subsection{\label{res_struct}Structural properties}

\begin{figure}[b]
\includegraphics[width=120mm]{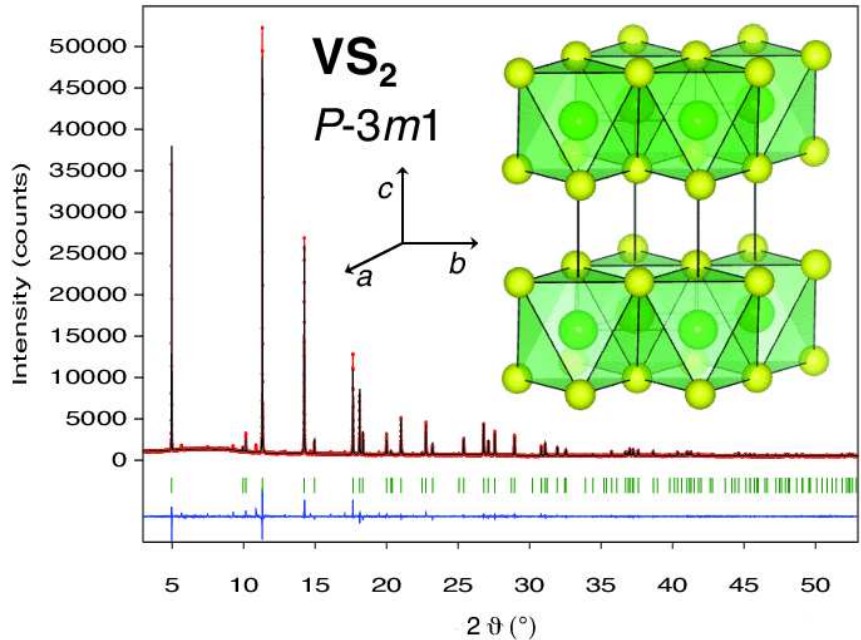}
\caption{\label{fig:XRD} (Color online) Synchrotron x-ray diffractogram measured on VS$_2$ powders at 5 K at the Materials Science beamline of the Swiss Light Source (SLS) using a wave length $\lambda$=0.49575 \AA. Red and black lines represent observed and calculated profile after Rietveld refinement, respectively, whilst the blue line is their difference. Green ticks represent the calculated position of the Bragg peaks (see also Table I). Inset: The 1T structure of VS$_2$ described by the $P\overline 3m$1 space group.}
\end{figure}

Rietveld refinements of the x-ray diffractograms measured on the as-prepared VS$_2$ powders confirm the previously reported 1T structure in the whole 5-270 K range and indicate that the samples are single-phase, with the exception of $\lesssim 0.3 \%$ vol. of PtS$_2$ impurity formed by the reaction of S with the Pt capsule. For all temperatures, good refinements with reliability factors $R_p \approx$2.5 were obtained in the $P\overline 3m$1 symmetry of the 1T structure by assuming no interstitial vanadium atoms ($x$=0). The result of the refinement is reported in Fig.~\ref{fig:XRD} and Table I for the 5 K data. The small thermal parameters indicate a good degree of crystallinity and a limited disorder. In the above symmetry, all the V-S bond distances in the VS$_6$ octahedron are equal and the only structural distortion allowed is a tilt of the octahedron axis with respect to the octahedron plane. We find a small tilt of 3.90(4)$^{\circ}$, which indicates that the octahedra are almost regular. We also verified the possibility of interstitial V atoms at the 1b site (0,0,1/2). For the 5 K data, this refinement yielded a site occupancy factor $x$=0.05, though the reliability factor was found to improve only slightly from $R_p$=2.53 to 2.20. We thus conclude that the excess $x$, if any, is less than 0.05.

\begin{figure}[b]
\includegraphics[width=110mm]{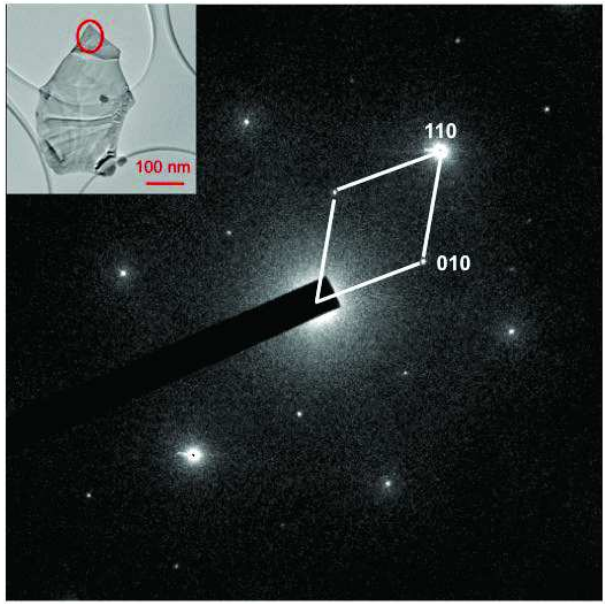}
\caption{\label{fig:TEM} Transmission Electron Microscopy diffraction pattern taken at 94 K on a grain of VS$_2$ marked by a circle in the inset. The ($hk$0) Bragg peaks and the in-plane $P\overline 3m$1 unit cell are shown. The absence of satellite peaks rule out the presence of long-range structural modulations.}
\end{figure}

\begin{figure}[b]
\includegraphics[width=88mm]{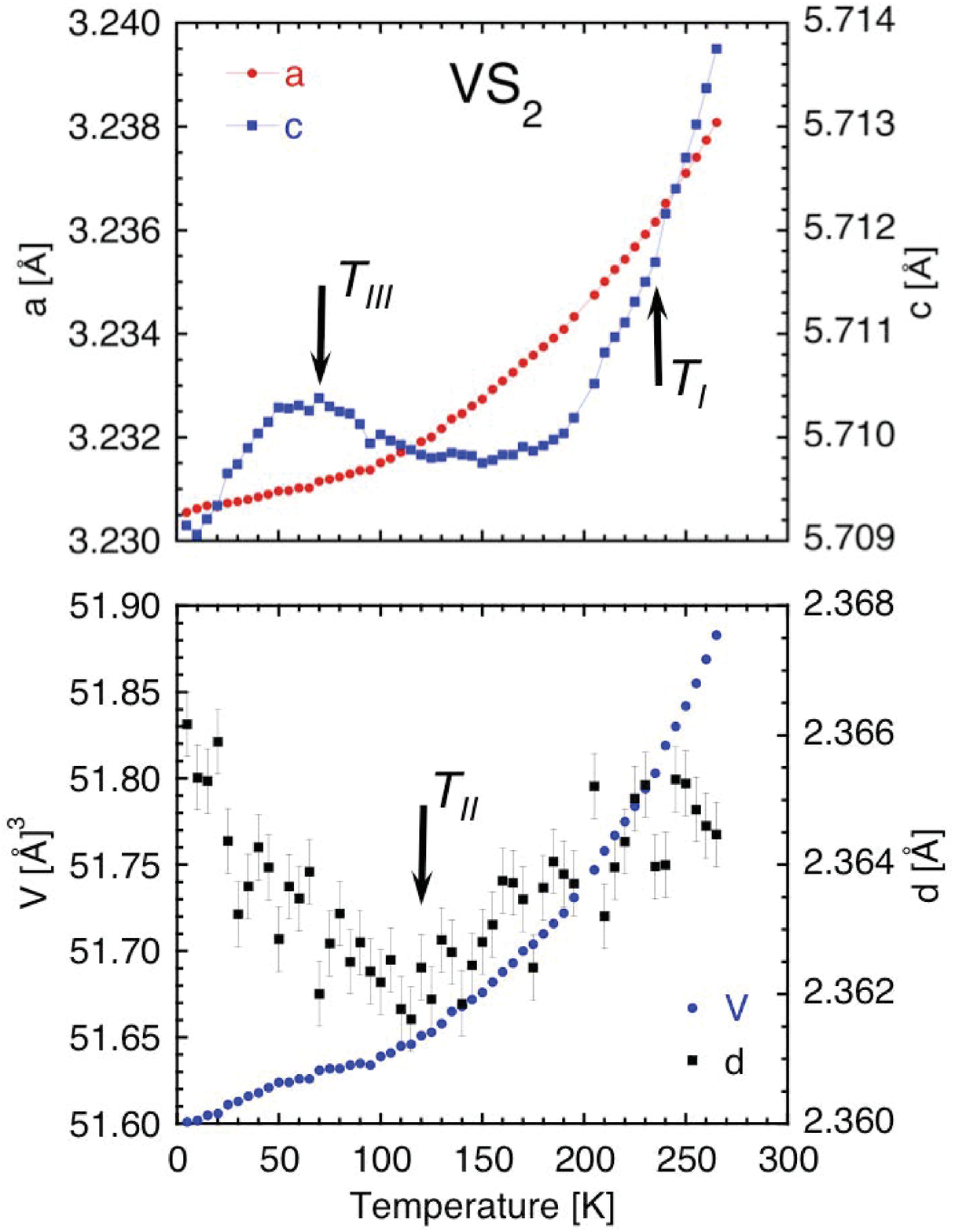}
\caption{\label{fig:latpar} Temperature dependence of the lattice parameters, $a$ and $c$ (top panel) and of the unit cell volume, $V$, and the V-S distance, $d$, (bottom) of VS$_2$ obtained from the Rietveld refinement of the powder diffraction data described in the text. The anomalies of the $c$-axis and of the V-S distance at $T_{II}$ and $T_{III}$ are discussed in the text. $T_{I}$ indicates the temperature at which the electrical resistivity of Fig.~\ref{fig:rho} exhibits a minimum.}
\end{figure}

Fig.~\ref{fig:latpar} summarizes the temperature dependence of the structural parameters. Whilst the behavior of the $a$-axis is normal, the $c$-axis exhibits an anomalous upturn at $T_{II} \approx 120$ K, which leads to a broad maximum at $T_{III} \approx 50$ K. This anomaly is associated with a V-shaped behavior of the V-S distance, $d$, showing a sudden linear increase with decreasing temperature at $T_{II} \approx 120$ K (see Fig.~\ref{fig:latpar}b). By extrapolating the high-temperature behavior of $d$ down to low temperatures, it is found that the anomaly corresponds to an expansion of $d$ by $\sim 0.01$ \AA. Notable is the fact that the $a$- ($c$-) axis parameter of the present samples is $\sim$1 \% longer (shorter) than those of the Li de-intercalated samples \cite{mur77}. Further evidence of a structural difference between the two types of samples is given by TEM. Contrary to a previous report on de-intercalated crystals \cite{mul10}, the present TEM diffraction patterns exhibit only the main spots of the hexagonal lattice and no trace of satellite peaks are found either at room temperature or at 94 K (see Fig.~\ref{fig:TEM}). In conclusion, no CDW phase or any other long-range structural modulations are detected, although the anomalous behavior of the V-S distance points at an incipient structural instability.

\begin{table}
\label{tab:structure}
\caption{Refined structure of VS$_2$ in the trigonal $P$-3$m$1 space group at 5 K. Refined lattice parameters are $a$=$b$=3.23055(1) \AA, $c$=5.70915(2) \AA. Numbers in parentheses indicate statistical uncertainty. Atomic coordinates $x$, $y$ and $z$ are in reduced lattice units.}
\begin{ruledtabular}
\begin{tabular}{ccccccc}

Atom & Wyckoff pos. & Site symmetry & $x$ & $y$ & $z$ & $B_{iso}$ ($\times 10^{-4}$ \AA$^2$)\\

\hline

V       & 1a & $m$ & 0 & 0  & 0 & (*)  \\
S       & 2d & 1   & 1/3 & 2/3 & 0.25503(12) & 0.326(11)  \\

\hline
\multicolumn{7}{l}
{(*)$B_{eq}$=0.920(18). Anisotropic $\beta_{ij}$ parameters ($\times 10^4$) for the V atom:}\\
\multicolumn{7}{l}{$\beta_{11}$=$\beta_{22}$=236.4(4.3); $\beta_{33}$=22.4(2.3); $\beta_{12}$=-118.2(2.1); $\beta_{13}$=0; $\beta_{23}$=0.}\\
\hline
\multicolumn{7}{l}{Reliability factors with all non-excluded points (not corrected for background):}\\
\multicolumn{7}{l}{$R_p$=2.53; $R_{wp}$=4.52; $R_{exp}$=3.40; $\chi^2$=1.76}\\

\end{tabular}
\end{ruledtabular}
\end{table}

\subsection{\label{res_mag}Magnetic properties}
\begin{figure}[b]
\includegraphics[width=88mm]{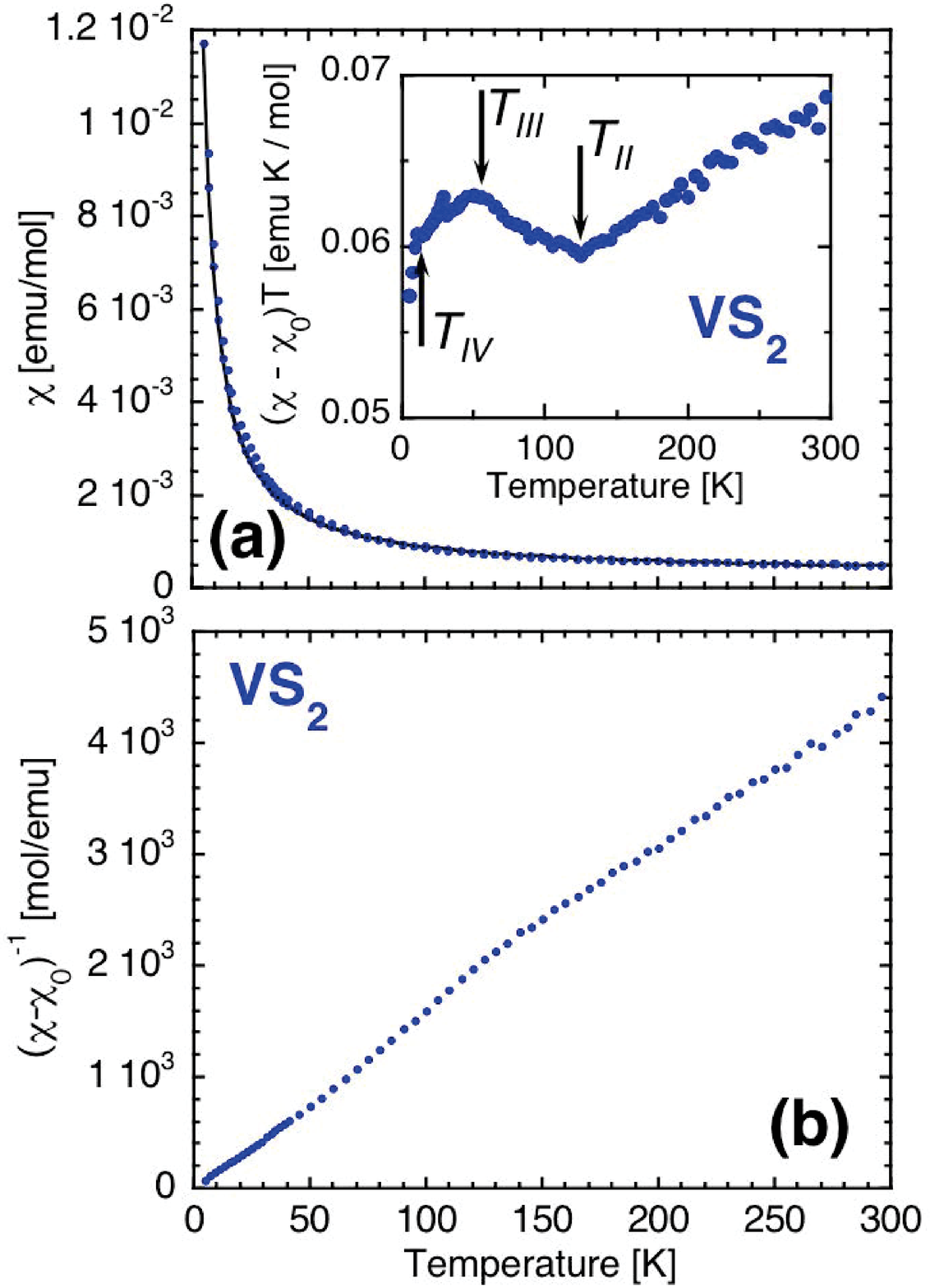}
\caption{\label{fig:susc} Top panel (a): temperature dependence of the magnetic susceptibility, $\chi(T)$, of a representative VS$_2$ sample measured at 10 oersted in both, zero-field- and field-cooling (ZFC, FC) mode. The two ZFC and FC curves are identical within the experimental error. The continuous line represents a Curie-Weiss fit of the ZFC data, as described in the text. Inset: the same data are plotted as $(\chi-\chi_0)T$ vs. $T$ in order to put into evidence the deviation of the data from the ideal Curie-Weiss dependence. The anomalies at $T_{II}$, $T_{III}$ and $T_{IV}$ are concomitant to the anomalies of Figs.~\ref{fig:latpar} and ~\ref{fig:rho} and discussed in the text. Bottom panel (b): temperature dependence of the inverse magnetic susceptibility, from which the constant term $\chi_0$ has been subtracted, which gives evidence of the overall Curie-Weiss behavior of the data with negligible Weiss constant, $\vartheta \approx 0$.}
\end{figure}

\subsubsection{Magnetic susceptibility measurements}
Fig.~\ref{fig:susc} shows the temperature dependence of the dc magnetic susceptibility, $\chi(T)$, measured at 10 oersted. The ZFC and FC curves are smooth and display no hysteresis, thus no magnetic orderings or other phase transitions occur, in agreement with the x-ray diffraction results. In Li de-intercalated VS$_2$ \cite{mul10} and in the isostructural and isoelectronic compounds VSe$_2$ \cite{dis81} and TaS$_2$ \cite{dis80}, the CDW transition manifests itself as an abrupt drop of $\chi$. The absence of this feature in our samples confirms the TEM evidence of no structural modulation. The $\chi(T)$ curve is well described by a conventional Curie-like term, $\chi(T) = C / (T + \vartheta)$, in addition to a constant paramagnetic term, $\chi_0$, with the following fitting parameters: $\chi_0 = 2.62\pm 0.03 \times 10^{-4}$ emu mol$^{-1}$, a Curie constant, $C = 6.29\pm 0.02 \times 10^{-2}$ emu mol$^{-1}$ K and a vanishing Weiss constant, $\vartheta = -0.46\pm 0.01$ K. These values are similar to those found in the isostructural and isoelectronic compound VSe$_2$ in the CDW phase \cite{dis81}.

Assuming a band picture and that the above $\chi_0$ value entirely arises from the Pauli contribution, we estimate a sizable density of states at the Fermi level, $g(E_F)=\chi_0/\mu_B^2 \approx$8.1 states eV$^{-1}$ cell$^{-1}$. However, the NMR data reported below indicate a significant orbital contribution, hence we believe that the above $g(E_F)$ value is overestimated and a more realistic estimate is given in the section devoted to the NMR results. Li de-intercalated VS$_2$ samples exhibit two times larger $\chi_0$ values $\approx 6.0 \times 10^{-4}$ emu/mol \cite{mur77,mul10}, so our samples are expected to be less metallic than the latter ones, in agreement with the optical absorption data presented below. The measured Curie constant, $C$, corresponds to an effective moment $\mu$=0.79 $\mu_B$ per V ion much smaller than the value $\mu = 1.73 \mu_B$ expected for $S=1/2$ V$^{4+}$ ions within a ionic picture. This reduction suggests a two-band scenario, where a majority of electrons are delocalized, whilst a minority $\approx$21\% form localized moments. 

A departure from the ideal Curie behavior is noted in the $(\chi -\chi_0)T$ vs. $T$ plot of Fig.~\ref{fig:susc}; three anomalies are seen at $T_{II}$, $T_{III}$ and $T_{IV}$. The first two ones are concomitant to the structural anomalies mentioned above and shown in Fig.~\ref{fig:latpar}. The $V$-shaped feature at $T_{II} \approx 120$ K mimics a similar feature in the temperature dependence of the V-S bond distance, whilst the maximum at $T_{III} \approx 50$ K follows the maximum of the $c$-axis lattice parameter in Fig.~\ref{fig:latpar}. These anomalies are also found in the NMR response and in the electrical resistivity (see below), so they should reflect subtle changes of the electronic structure.  

\subsubsection{NMR measurements}
In the whole 1.6 - 300 K range studied, we measured a strong $^{51}$V NMR signal. Note that this signal does not arise from the localized moments of the V ions because the nuclei of paramagnetic V$^{4+}$ ions probe a large fluctuating hyperfine field $B_{hf} \approx$ 10 T \cite{freeman_watson,oha99,kik94}. This field causes exceedingly fast nuclear relaxations that can not be detected, unless a strong exchange interaction narrows the resonance line. This is not the case here, considering the negligible exchange energy measured, $J/k_B \sim \vartheta \lesssim$ 1 K. We estimated the amplitude and correlation time of the magnetic fluctuations at the nucleus to be $\delta \omega={^{51}\gamma} B_{hf} \approx 7\times 10^8$~s$^{-1}$ and $\tau=\hbar/J \approx 10^{-11}$ s, respectively, where $^{51}\gamma/2\pi = 11.19$ MHz/T is the nuclear gyromagnetic ratio of $^{51}$V. From these values, one obtains a transverse relaxation time $T_2\approx (\delta\omega^2\,\tau)^{-1}\approx$ 0.2 $\mu$s much shorter than the instrumental dead time, $\sim$10 $\mu$s \cite{abragam}. Thus, the nuclear resonance of the $V^{4+}$ ions is not detectable and the observed signal originates from non-magnetic V.

\begin{figure}[b]
\includegraphics[width=88mm]{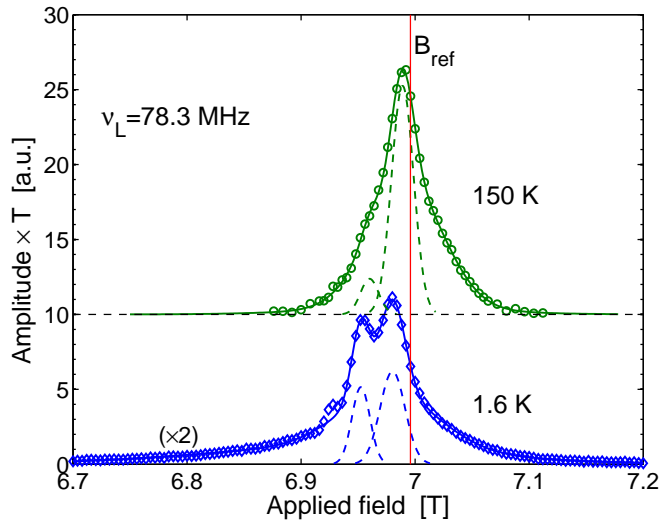}
\caption{\label{fig:NMR} (Color online) $^{51}$V field-sweep NMR of VS$_2$ at 1.6 and 150 K, recorded at 78.3 MHz. The reference field, $B_{ref}$, is indicated by the vertical red line. For clarity, the 1.6 K spectrum has been scaled by a factor of 2 after correction for the nuclear susceptibility $\propto 1/k_BT$. The continuous line is a fit of the the two sharp peaks indicated by the broken line. The small peak at 6.927 T produced by the $^{63}$Cu signal of the copper coil has been excluded from the fit.}
\end{figure}

In Fig.~\ref{fig:NMR} we show two representative $^{51}$V field-sweep spectra at 150 and 1.6 K. They are characterized by two sharp and inhomogeneously broadened peaks and by a broad shoulder. Since the sample is single-phase, this indicates that the non magnetic V ions probe two distinct electronic environments, which supports an electronic phase separation picture in the local scale. Both peaks exhibit a sizable positive Knight shift of the resonance line with respect to the reference field $B_{ref}$=6.9959 T. Upon cooling, the two-peak structure is better resolved due to a transfer of spectral weight from the the main peak at higher field to the minor peak. The spectra do not exhibit the characteristic quadrupolar pattern of $I = 7/2$ nuclear spins expected for powder samples of $^{51}$V. This pattern would display a sharp central line and $2I-1$ satellite lines shifted by the quadrupolar field and broadened by the random orientation of the electric field gradient (EFG), which gives rise to peculiar powder singularities. We believe that the quadrupolar structure is not resolved because the high fields used cause a large magnetic broadening of a spatially inhomogeneous distribution of Knight shifts. This explanation is supported by a former estimate of the quadrupolar frequency, $\nu_Q$. Specifically, for VS$_2$ \cite{tsu83} and VSe$_2$ \cite{tsu81}, a value $2\pi \nu_Q/^{51}\gamma \approx$ 0.033 T (in field units) was found. Since $\nu_Q$ is proportional to the EFG at the nucleus, the broad shoulders are attributed to the quadrupole satellite transitions in presence of magnetic and possibly EFG inhomogeneities, which would smear the powder singularities.

\begin{figure}[b]
\includegraphics[width=88mm]{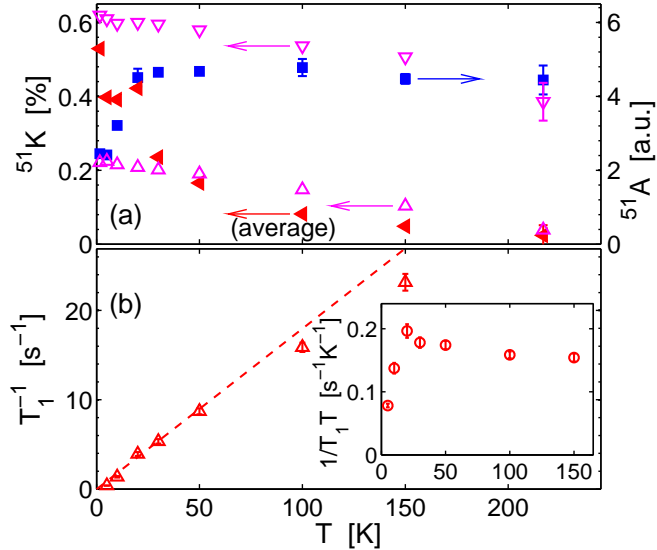}
\caption{\label{fig:Knight} (Color online) a) Temperature dependence of the Knight shifts $^{51}K$ of the two peaks (open triangles), the center of gravity of the spectra $^{51}\overline K$ (filled triangles) and the integrated spectral amplitude multiplied by temperature $^{51}A$ (squares). b) Temperature dependence of the spin-lattice relaxation rate $T_1^{-1}$ and best fit to the Korringa law $T_1^{-1} \propto T$ (dashed line). Inset: the same as before for the Korringa product $(T\,T_1)^{-1}$.}
\end{figure}

The two sharp peaks are fitted well in the whole 1.6-300 K range studied, which enables us to plot the Knight shifts $^{51}K$ of the two peaks and the shift $^{51}\overline{K}$ of the center of gravity of the spectrum as a function of temperature (see Fig.~\ref{fig:Knight}a). For both peaks, $^{51}K$ increases upon cooling down to $T_{III} \approx$ 50 K, where an anomaly in the susceptibility data is also observed, then it levels off. Below $T_{II} \approx$ 100 K, the average shift $^{51}\overline{K}$ exhibits a more marked temperature dependence than that of the single peaks owing to the aforementioned transfer of spectral weight. The plot also shows the temperature dependence of the NMR signal amplitude, $^{51}A$, integrated over the spectrum and corrected for the nuclear susceptibility $\propto 1/k_BT$. Note that $^{51}A$ is nearly $T$-independent at high temperature, whilst it drops by $\approx$ 50\% below $T_{IV} \approx$ 20 K, where the susceptibility also decreases. Considering that $^{51}A$ is directly proportional to the number of $^{51}V$ nuclei detected in the resonance, this loss of signal indicates that, for a sizable fraction of the nuclei, the spin-spin relaxation time, $T_2$, is significantly reduced. This prevents detection if this time becomes shorter than the dead time of the NMR receiver. It follows that the correlation time of the fluctuations probed by these nuclei increases dramatically below $T_{IV}$. The nature of such slow fluctuations remains to be established; they may be either magnetic or electric, \textit{e.g.} due to slow charge motion, which would produce random EFG modulations. The $^{51}K$ data provide further evidence that the NMR signal does not arise from localized V moments. Namely, the orbital contribution to $K$ and to the susceptibility $\chi$ is usually negligible for a magnetic ion, thus $K$ is proportional to $\chi$ via the hyperfine coupling constant and $K$ should follow the Curie law, which is not observed here. Moreover, in a magnetic ion, a \textit{negative} hyperfine field at the nucleus is expected \cite{freeman_watson}, in contrast with our observation of \textit{positive} $K$. 

The temperature dependence of the spin-lattice relaxation was measured for the main peak. The recovery of the nuclear magnetization $M(t)$ to thermal equilibrium was found to obey the following multiexponential law, which is appropriate for the magnetic relaxation of the quadrupole-resolved central line of a $I=7/2$ nucleus following its selective saturation: \cite{narath}
 
\begin{equation}
\label{eq:SR7_2}
\frac{M(t)}{M(t=\infty)} = 1 - \frac{4}{21}e^{-2Wt} -
\frac{2}{11}e^{-12Wt} - \frac{20}{91}e^{-30Wt} - \frac{175}{429}e^{-56Wt} 
\end{equation}

where $W$ is the transition probability between two adjacent nuclear Zeeman levels. In Fig.~\ref{fig:Knight}b, the spin-lattice relaxation rate $T_1^{-1}\equiv 2W$, determined by the analysis of $M(t)$ using Eq.~\ref{eq:SR7_2}, is plotted as a function of temperature. In the 10-150 K range, the data follow the Korringa law $T^{-1}_1 \propto T$, characteristic of a non-magnetic metal with a temperature-independent spin susceptibility. The validity of this law in the above range is also apparent in the $(T\,T_1)^{-1}$ vs.\ $T$ plot, which shows a nearly constant value $\approx 0.17$~K$^{-1}$s$^{-1}$. At lower temperature below $T_{IV}$, a large deviation from this simple behavior is found.

In a metal, $K$ and $T_1^{-1}$ probe the static and dynamic electronic spin susceptibilities, respectively. The two quantities are then predicted to scale with each other according to the universal Korringa relation: 

\begin{equation}
\label{eq:korringa}
K^2 = R\, (T\,T_1)^{-1}, 
\end{equation}

where $R$ is the Korringa ratio. For a simple $s$-band metal with negligible electronic correlations, $R$ is expected to be equal to the universal value $S_0 = (\hbar/4\pi k_B) (\gamma_e/\gamma_n)^2$, where $\gamma_e$ and $\gamma_n$ are the electronic and nuclear gyromagnetic ratios \cite{abragam}. For a $d$-band metal, a similar result holds, with $R=\kappa S_0$ and $2 \le \kappa \le 5$.\cite{yafet_jaccarino} From Fig.~\ref{fig:Knight}, it is clear that the above relation can not be followed by our temperature-dependent $^{51}K$ and temperature-independent $(T\,T_1)^{-1}$ data. Even restricting ourselves to the 10 K $\le T \le$ 50 K range, where both quantities are approximately constant, anomalously high values $R/S_0=10-100$ are obtained. This is explained by comparing our data with those by Tsuda {\it et al.} \cite{tsu83} who found one order of magnitude smaller Knight shifts but a comparable $(T\,T_1)^{-1}\approx 0.4$ K$^{-1}$s$^{-1}$ at $T<$100 K, implying a spin susceptibility smaller by a factor of 1.5 only in our sample. The difference between the $T_1$ values reported by Tsuda \textit{et al.} and ours is not significant, for the recovery law of a quadrupolar nucleus depends upon the experimental conditions, \textit{e.g.} the width of the irradiated band and the length of the saturation pulse train \cite{and61,reg91}. This may lead to deviations from Eq.~\ref{eq:SR7_2}, whence an uncertainty on $T_1$. Tsuda \textit{et al.} showed that $^{51}K$ contains two contributions with opposite sign which nearly cancel each other: a positive orbital one, $^{51}K_{orb}$, and a negative one, $^{51}K_d$, proportional to the spin susceptibility of the $d$-wave conduction band. The discrepancy between Tsuda's $^{51}K$ values and ours are then reconciled by assuming a similar value of the $K_d$ term (and a similar Korringa product) for the two samples but a much larger orbital contribution in our sample, which explains the larger Korringa ratios $R$. A large orbital shift $^{51}K_{orb}$ in $d$-wave metals originates from the presence of low-lying excited multiplets via the van Vleck mechanism \cite{clo64, wal08} and thus depends upon the crystal field splitting. Since the latter is sensitive to the local structural parameters, we believe that the anomaly of $^{51}K$ and $^{51}A$ at $T_{III}$ reflects the corresponding anomaly of the $c/a$ ratio and of the V-S bond distance.

To conclude this section, we endeavor to estimate the Pauli contribution $\chi_d$ to the magnetic susceptibility from the $T_1^{-1}$ data and from the values of the Korringa and hyperfine constants determined experimentally in previous reports. According to band calculations\cite{myron80} and ARPES data \cite{mul10}, we should assume a full $d$-wave character of the conduction band. The Korringa ratio was determined to be $R=5.0\times 10^{-6}$~sK in both, VS$_2$ and VSe$_2$ \cite{tsu83,tsu81}. According to Eq.~\ref{eq:korringa}, we obtain $K_d=-0.09$~\% for our sample. Next, we consider the proportionality relation $^{51}K_d={\cal A}_{iso}\chi_d /N_A$, where ${\cal A}_{iso}$ is the isotropic core-polarization term in the hyperfine coupling of vanadium and $N_A$ is the Avogadro number. From the electron spin resonance of the V$^{3+}$ ion \cite{epr96,epr94}, we take ${\cal A}_{iso}=-85$~kOe/$\mu_B$ to obtain $\chi_d = 6\times 10^{-5}$ emu/mol, which corresponds to a density of states $g(E_F) \approx$ 2.0 states eV$^{-1}$ cell$^{-1}$. We finally obtain $\chi_{orb}=\chi_0-\chi_d\approx2\times 10^{-4}$~emu/mol, comparable with the values reported in the related compound LiVO$_2$ \cite{Vorbital}.

\begin{figure}[b]
\includegraphics[width=96mm]{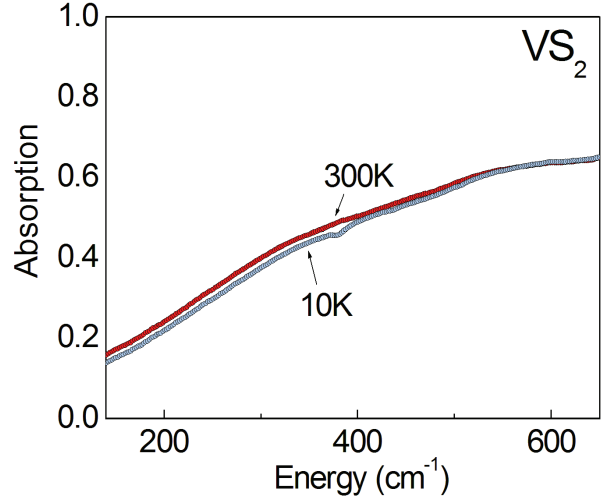}
\caption{\label{fig:opt_cond} Temperature dependence of the optical absorption of pure VS$_2$ powders as a function of frequency in the 4-300 K range. The absence of a Drude peak and a weak increase of absorption with temperature indicate a nonmetallic behavior. Note the absence of phonon peaks, except for a weak absorption structure at $\approx$390 cm$^{-1}$ ($\approx$ 49 meV) suggestive of a highly screened phonon mode, in agreement with the prediction of a $E_u$ infrared-active mode at 405.8 cm$^{-1}$ (see text).}
\end{figure}

\subsection{Transport properties}
\subsubsection{Infrared optical conductivity}
The optical absorption spectra, $A_{\omega}$, measured in the infrared 140-650 cm$^{-1}$ range is shown for two representative temperatures, 5 and 300 K, in Fig.~\ref{fig:opt_cond}. Salient features are: (i) the absence of a Drude peak; (ii) a sizable and monotonic increase of $A_{\omega}$ with $\omega$; (iii) the absence of phonon features at high temperature and one broad phonon feature at $\omega_{ph} \approx 390$ cm$^{-1}$ at low temperature; (iv) the spectra change little with temperature in the whole 5-300 K range. Specifically, besides the appearance of the above phonon feature at low temperature, the only noticeable change is a small increase of the absorption with temperature at low frequency. Considering that, under the present experimental conditions of low absorption, $A_{\omega}$ is simply proportional to the real part of the optical conductivity, $\sigma_{\omega}$, features (i) and (ii) give evidence of non-metallic transport, whilst (iv) suggests a modest charge localization. According to a symmetry analysis of the 1T structure, feature (iii) is consistent with the expectation of two infrared-active phonon modes with $A_{2u}$ and $E_u$ symmetry. It is then plausible that only one of these modes is detected at low temperature; indeed, the measured frequency, $\omega_{ph}$, is in good agreement with the calculated value of 405.8 cm$^{-1}$ for the $E_u$ mode (see section on phonon calculations). The fact that the phonon feature at $\omega_{ph}$ is visible only at low temperature and is broadened indicates a partial screening of the infrared active dipoles, in agreement with the observation of Korringa behavior in the NMR data and of a significant density of states extracted from the measured $\chi_0$ value and calculated \textit{ab initio} (see below). On the other hand, the absence of a Drude peak indicates that the electron density responsible for the screening is not sufficient to ensure a metallic transport, possibly because of disorder or electron-electron correlations leading to charge localization. Conversely, a picture of conventional metal and semimetal accounts for the infrared response of other transition metal dichalcogenides, such as 2H-TaS$_2$ \cite{hu07} and 1T-TiSe$_2$ \cite{li07}, respectively, which both exhibit well-developed Drude peaks.

A picture of charge localization has been previously proposed also for the isostructural and isoelectronic $d^1$ compound 1T-TaS$_2$ \cite{faz80,sip08}, which indeed displays a qualitatively similar optical response in the so-called nearly commensurate CDW (NCCDW) phase at room temperature \cite{gas02}. On the other hand, in the commensurate CDW phase at low temperature, the optical spectra of 1T-TaS$_2$ \cite{gas02} and also of 1T-TiSe$_2$ \cite{li07} display a large number of sharp phonon features, consistent with the large number of infrared phonon modes expected for the CDW superstructure. The absence of such additional phonon features in our samples is a further strong argument against the existence of a CDW phase. To the best of our knowledge, the present optical data are the first ones reported in VS$_2$, so no direct comparison can be made with Li de-intercalated samples; it would be interesting to see whether these samples display additional infrared phonon features in the CDW phase. 

\subsubsection{Electrical resistivity}
\begin{figure}[b]
\includegraphics[width=96mm]{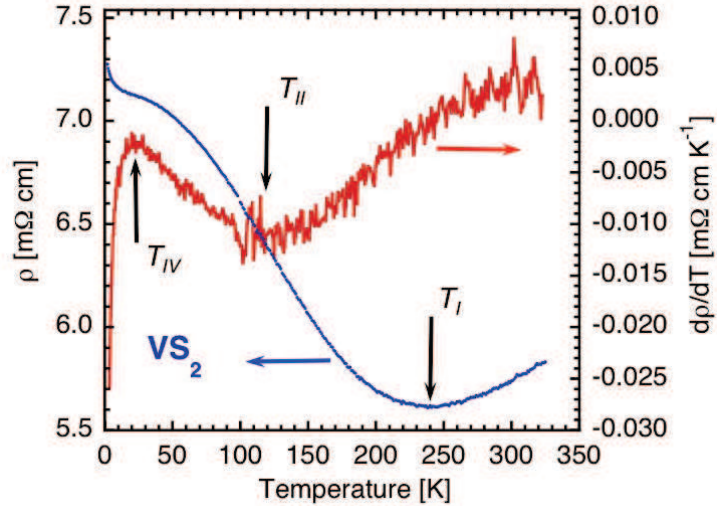}
\caption{\label{fig:rho} Temperature dependence of the dc electrical resistivity, $\varrho$, and of its derivative. Note the characteristic semimetallic behavior characterized by a crossover from a positive to a negative slope of $\varrho$ at $T_I$, and two abrupt changes of the slope at $T_{II}$ and $T_{IV}$ concomitant to anomalies of the lattice parameters and of the magnetic susceptibility (see Figs. ~\ref{fig:latpar},\ref{fig:susc}).}
\end{figure}

In Fig.~\ref{fig:rho}, we report the electrical resistivity curve $\varrho (T)$ measured in the 2-325 K range on one representative VS$_2$ sample. Consistently with the optical conductivity results, the overall temperature dependence of $\varrho$ exhibits a modest variation within the 5.5-7.2 m$\Omega$ cm range. Specifically, at low temperature, one notes a weak increase which confirms the above evidence for charge localization. At high temperature, the weakly negative resistivity coefficient above $T_{I} \approx$ 250 K is explained by a small increase of thermally activated carriers, which is characteristic of a small energy gap at $E_F$. A comparison between the present resistivity data and those obtained on Li de-intercalated samples \cite{mul10} further confirms that the latter samples are different from the present ones. First, we find no anomaly of $\varrho(T)$ at $T_{CDW}$=305 K, which supports the conclusion on the absence of CDW transition. Second, the marked metallic behavior of the above samples below $\approx$260 K, with a sizable residual resistivity ratio, $RRR \sim 10$, is completely different from the nonmetallic behavior reported here. On the other hand, the present data display two anomalies at $T_{III} \approx 120$ K and $T_{IV}\approx 20$ K concomitant to the anomalies in the dependence of the V-S distance and of the magnetic susceptibility with temperature shown in Fig.~\ref{fig:latpar} and Fig.~\ref{fig:susc}, respectively. In Fig.~\ref{fig:rho}, these anomalies appear as inflexion points of the $\varrho (T)$ curve and are better seen in the derivative of the curve. The inflexion point at $T_{III}$ corresponds to a slowing down of the resistivity increase upon cooling, which is interpreted as a weakening of the localization below $T_{III}$. The opposite seems to occur at $T_{IV}$. In order to provide a full picture of the transport properties of the present high-pressure VS$_2$ phase and to confirm the proposed scenario of charge localization, a single crystal study would be desirable.    

\subsection{Electronic structure and lattice stability}

We first performed a full structural optimization of the crystal structure by varying both, internal coordinates and unit cell parameters. We found that the LDA result substantially underestimates the experimental volume. Namely, we obtained $a_{\rm LDA}=3.102$ ~\AA, $c_{\rm LDA}/a_{\rm LDA} = 1.786$ and a $z$ coordinate of $0.262$ (in reduced coordinates) for the $S$ atom. Since a compressed volume tends to weaken the tendency towards CDW formation (it is recalled that the CDW phase disappears under hydrostatic pressure in all metallic transition metal dichalcogenides), here we used the experimental lattice parameters and optimized only the internal coordinate of the $S$ atom in order to investigate the stability of the structure towards CDW formation. 

\subsubsection{Electronic band structure}
The electronic structure of VS$_2$ shown in Fig.~\ref{fig:bands} turns out to be very weakly dependent on the volume used in the calculations. The overall shape of the band-structure closely recalls that of TiSe$_2$ \cite{cal11} but with a different position of the Fermi level and a smaller hybridization between the chalcogen $p$ states and the transition metal $3d$ states. According to the band structure, VS$_2$ should be a metal with a density of states at the Fermi level $g(E_F)$ = 3.2 states eV$^{-1}$ cell$^{-1}$. This value is in agreement with the $\chi_0$ value obtained from susceptibility measurements, considering that only one third of this value arises from the Pauli term, as suggested by the analysis of the NMR data. As to the band characteristics at the Fermi level, the $d$-electrons form electron-pockets at the $M$ and $L$ symmetry points and an additional small electron-pocket is present along the $K-A$ direction.

\subsubsection{Phonon dispersion}
\begin{figure}[b]
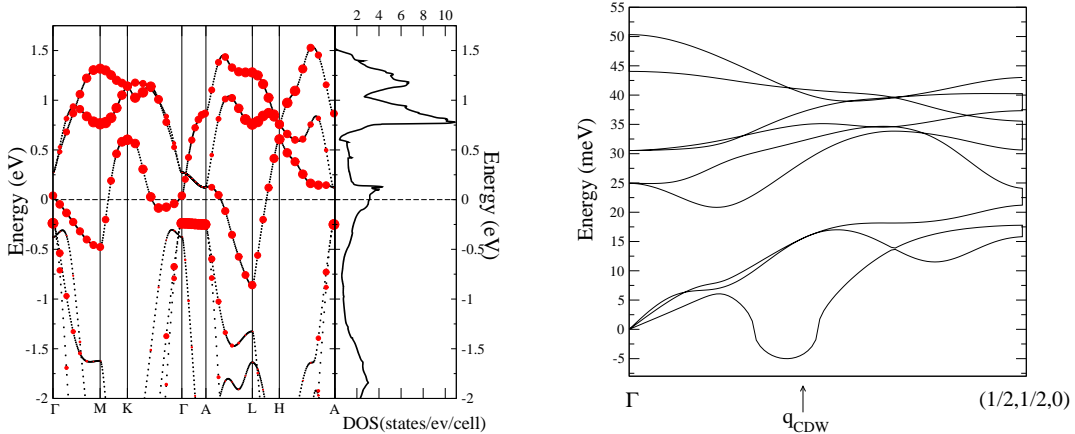

\includegraphics[width=0.4\columnwidth]{VS2_electron_bands.eps}\hspace{1.0cm}\includegraphics[width=0.4\columnwidth]{VS2_phonon_bands.eps}
\caption{\label{fig:bands}Left: Electronic band structure and density of states of VS$_2$ in the local density approximation. The size of the circles on a given band is proportional to the vanadium 3$d$-component of the band. Right: Phonon dispersion along the $(1,1,0)$ reciprocal space direction. The label $q_{\rm CDW}$ indicates the propagation vector of the CDW instability reported in Ref. \onlinecite{mul10}.}
\end{figure}

An insight into the occurrence of a CDW transition is obtained from the harmonic phonon dispersion calculation reported in Fig.~\ref{fig:bands}. Within the harmonic approximation, an imaginary phonon frequency (here plotted as negative) is the signature of a second order structural instability, such as a CDW. In VS$_2$, we do indeed find that the harmonic phonon dispersion shows an instability of a transverse acoustic phonon at the same wave vector ${\bf q}_{CDW}$=(0.21,0.21,0) of the CDW phase found experimentally in Ref. \cite{mul10}. The \textit{caveat} is that the calculation turns out to be extremely sensitive to the values of lattice parameters (much more than in the case of other dichalcogenides) and the unstable phonon mode is only slightly imaginary. This suggests that the structure can be stabilized by anharmonic effects. This is what indeed occurs in the parent compound 2H-NbS$_2$ \cite{PhysRevB.86.155125}, where harmonic calculations overestimates the tendency towards CDW formation and anharmonic effects stabilize the lattice without CDW. The lack of anharmonic effects in the present calculations and the high sensitivity of the DFT phonon dispersion on the lattice parameters do not allow us to draw more definitive conclusions. It is nevertheless safe to conclude that VS$_2$ is at the verge of a CDW instability, which accounts for the contrasting results about the occurrence of a CDW phase in Li de-intercalated samples and in the present ones. Finally, the calculations predict two infrared-active $E_u$ modes at 200.9 cm$^{-1}$ and 405.8 cm$^{-1}$ and two Raman-active ones at 246.3 cm$^{-1}$ and 355.3 cm$^{-1}$ with $E_g$ and $A_{1g}$ symmetry, respectively. The prediction of the 405.8 cm$^{-1}$ mode is consistent with the observation of a highly screened phonon mode at a similar frequency in the optical absorption, as discussed above.    
  
\section{\label{concl}Discussion and conclusions}
In conclusion, we systematically studied the structural, magnetic and transport properties of single-phase 1T-VS$_2$ samples prepared under high pressure. Contrary to previous reports on Li de-intercalated samples, we found that 1T-VS$_2$ is nonmetallic and displays no long-range structural modulations, such as CDW. This difference is attributed to the different synthesis route employed which does not require any chemical methods. Specifically, we envisage that residual Li atoms and iodine or acetonitrile molecules introduced by the Li de-intercalation method may alter the doping level, which would be sufficient to stabilize the metallic and CDW phase owing to the semi-metallic character of the band structure. To the best of our knowledge, the doping effect inherent in the de-intercalation process has not been considered in previous studies but is supported by the observation of structural modulations in chemically exfoliated MoS$_2$ and WS$_2$ by means of Li intercalation \cite{eda12}. This scenario would also explain the different lattice parameters of the present VS$_2$ samples with respect to those of the Li de-intercalated ones. A further support of this scenario is given by the enhancement of metallic properties in the related layered compounds AuVS$_2$ \cite{gau02} and Ag$_{0.75}$VS$_2$ \cite{ali13}, where the nominal valence of V is close to $3+$ instead of $4+$. 

Within a rigid band picture, the evidence of nonmetallic conductivity would be at odds with the Korringa-like behavior of the spin-lattice relaxation time, the sizable Pauli susceptibility, $\chi_0 \sim 10^{-4}$ emu mol$^{-1}$, and the prediction of metallic properties by band structure calculations. This discrepancy indicates that the above picture is not appropriate and localization effects due to disorder or electronic correlations should be invoked. An enhancement of the Pauli susceptibility induced by these effects in quasi two-dimensions would also account for the comparable $\chi_0$ values reported in other 1T-VS$_2$-related compounds, regardless to their metallic or insulating properties. Examples are the misfit layer system La$_{1.17-x}$Sr$_x$VS$_{3.17}$ \cite{yas95,car99}, which displays no appreciable change of $\chi_0$ across the doping-induced metal-insulator transition, the $M$V$_2$S$_4$ family ($M$=Ti, Cr, Fe, Ni or Cu) \cite{pow99,kle11}, where both metallic and insulating properties are found, and the insulator Sr$_3$V$_5$S$_{11}$\cite{kle13}. 

In support of this scenario, it is recalled that a Mott state has been proposed by several groups for 1T-TaS$_2$, which displays similar magnetic and transport properties in the aforementioned NCCDW phase  \cite{gee72,faz80,ros06,sip08}. In this phase, the direct STM observation of a nm-size domain structure \cite{bur91}, in excellent agreement with calculations \cite{nak84}, in conjunction with high-pressure resistivity measurements suggests a picture of insulating and commensurate CDW regions separated by metallic and weakly CDW-distorted ones \cite{sip08}. A phase separation scenario may then apply to the present case as well, for our magnetic susceptibility data unveil the existence of both localized moments and metallic carriers, whilst the NMR data show that the metallic V sites are inhomogeneous. The existence of nm-size domains would indeed explain the absence of satellite diffraction peaks in our TEM pictures, whilst a slightly different doping level in Li de-intercalated samples would be sufficient to stabilize a uniform metallic/CDW phase, in agreement with the prediction of a latent CDW instability by phonon calculations. The confirmation of a phase separation scenario for the present 1T-VS$_2$ samples awaits further studies by means of local probes, such as Scanning Tunneling Microscopy/Spectroscopy. 

The authors acknowledge L. Cario and M. Marezio for stimulating discussions and R. Lobo and B. L\'eridon for making available their PPMS system. MC gratefully acknowledges financial support of the Graphene Flagship and of the French National ANR funds under reference ANR-11-IDEX-0004-02, ANR-11-BS04-0019 and ANR-13-IS10-0003-01. Computer facilities were provided by CINES, CCRT and IDRIS (project no. x2014091202).

\bibliography{VS2_bib}

\end{document}